\documentclass[prd,a4paper,twocolumn]{revtex4} 

\usepackage{graphicx} 
\usepackage{subfigure,amssymb,latexsym} 
\usepackage{psfrag} 

\newcommand{\nc}{\newcommand} 

\nc{\be}[1]{\begin{equation}\mbox{$\label{#1}$}} 
\nc{\bea}[1]{\begin{eqnarray} \mbox{$\label{#1}$}} 
\nc{\eea}{\end{eqnarray}} 
\nc{\ee}{\end{equation}} 

\nc{\bc}{\begin{center}} 
\nc{\ec}{\end{center}} 
\nc{\ba}{\begin{array}} 
\nc{\ea}{\end{array}} 
\nc{\bab}{\begin{abstract}} 
\nc{\eab}{\end{abstract}} 
\nc{\btab}{\begin{tabular}} 
\nc{\etab}{\end{tabular}} 
\nc{\bit}{\begin{itemize}} 
\nc{\eit}{\end{itemize}} 
\nc{\ben}{\begin{enumerate}} 
\nc{\een}{\end{enumerate}} 
\nc{\bfig}{\begin{figure}} 
\nc{\efig}{\end{figure}} 

\nc{\dd}[2]{{{\partial #1}\over{\partial #2}}} 
\nc{\ddd}[2]{{{{\partial}^2 #1}\over{\partial {#2}^2}}} 
\nc{\dddd}[3]{{{{\partial}^2 #1}\over 
   {\partial #2 \partial #3}}} 

\nc{\ie}{{\it i.e. }} 
\nc{\eg}{{\it e.g. }} 

\nc{\mn}{{\mu\nu}} 

\nc{\gppn}{$\gamma_{PPN} \,$} 
\nc{\lcdm}{$\Lambda$CDM\,} 
\nc{\f}{\frac}

\begin{document}

\title{Is scalar-tensor gravity consistent with polytropic stellar models?}

\author{K. Henttunen} 
\email{kajohe@utu.fi} 
\author{I. Vilja} 
\email{vilja@utu.fi} 
\affiliation{Department of Physics and Astronomy, University of Turku, FIN-20014 Turku, FINLAND} 

\date{\today} 

\begin{abstract} 
We study the scalar field potential $V(\phi)$ in the scalar-tensor gravity 
with self-consistent polytropic stellar configurations. Without choosing a particular potential, 
we numerically derive the potential inside various stellar objects.
We restrict the potential to conform to general relativity or to $f(R)$ gravity inside and require the
solution to arrive at SdS vacuum at the surface.
The studied objects are required to obtain observationally valid masses and radii 
corresponding to solar type stars, white dwarfs and neutron stars.
We find that the resulting  scalar-tensor potential $V(\phi)$ for the numerically derived polytrope
that conforms to general relativity, in each object class, is highly dependent on the matter configuration
as well as on the vacuum requirement at the boundary. As a result, every stellar configuration
arrives at a potential $V(\phi)$ that is not consistent with the  other stellar class potentials.
Therefore, a general potential
that conforms to all these polytropic  stellar classes could not be found.
\end{abstract} 

\maketitle 

\section{Introduction}
General relativity (GR), describes the local gravitational phenomena
very well \cite{GRtests}. 
At large scales and in the early universe this description does not seem to be adequate anymore. 
From the observations of distant supernovae and cosmic microwave background \cite{sne1a,cmb} 
the expansion of the universe is interpreted to be accelerated at late times.
To encompass the late time accelerated phase GR needs to be modified. The simplest modification
allowed by the Einstein-Hilbert action is the cosmological constant model with cold dark matter, 
called the \lcdm model \cite{lcdm}.
In this work the \lcdm model is denoted as GR$+\Lambda$.
While \lcdm is very successful,
this model has its shortcomings as well \cite{DEmodels} and many ways to explain the current 
accelerated phase have been suggested. 
Models generally either modify the content of the Einstein's equations, by including
a dark energy component \cite{DEmodels},
or by modifying the gravity sector itself \cite{modgravcombo,0707.2748,FM03}.
Because of the great success of GR in predicting the local observations with high accuracy,
viable modifications must allow only configurations with small deviations from the 
general relativistic solutions.

A class of GR modifications that does not require any extra assumptions in addition to standard physics
is the $f(R)$ theories of gravity. In $f(R)$ theories of gravity, an algebraic function of the Ricci 
scalar extends the gravitational Lagrangian density from GR \cite{SotFar,fRths}.
Another way of obtaining adequate modifications to GR are so called scalar-tensor theories \cite{scal-tens}.
These scalar-tensor theories 
arise naturally from {\it e.g.} higher dimensions, string theory or from 
non-commutative geometry \cite{FM03}.
In the Jordan frame formulation of the theory the scalar field is non-minimally coupled to the metric tensor
but does not couple to the matter sector.
The scalar field, with an adequate potential,  can  give rise to the observed exponential acceleration 
today with a slow-roll behavior \cite{quint} and still conform to the local gravitational experiments 
via a chameleon mechanism \cite{chameleon}.
There are viable alternatives among the gravity modifications that do not possess instabilities \cite{instab,fRsolsyst},  
can account for the correct expansion history of the universe \cite{epochs} and even produce inflation in some cases \cite{fRinfl}. 
The local tests
that GR passes must also be accounted for \cite{fRsolsyst,GRSStests}. Some $f(R)$ models
can resemble GR so much that even the spacetime outside the $f(R)$-sun conforms to the observations 
\cite{HV14}.
This is, however, not a general feature of $f(R)$ theories of which most are excluded by the light 
deflection experiments \cite{cassini,Faraoni0610734}.
It is also known that the Birkhoff's theorem \cite{Birkhoff} is broken for the general relativity 
modifications \cite{Faraoni1001.2287}.  According to the Birkhoff's theorem, the vacuum field equations
of general relativity obtain the Schwarzschild solution around a spherically symmetric object.

In this work, 
we are looking for stellar solutions in the Jordan frame formulation of a scalar-tensor theory 
with GR$+\Lambda$ behavior inside the star.
We also demand the configurations to arrive at the Schwarzschild-de Sitter (SdS) vacuum at the boundary. 
The  polytropic stellar object classes:
solar type main sequence stars (SUN), white dwarfs (WD) and neutron stars (NS);
are the matter configurations for which the scalar field potential $V(\phi)$ was
solved numerically.
We find that a potential $V(\phi)$ that will describe all the studied object classes cannot be found.
The solution for the potential is found to be specific to each stellar class and
highly dependent on the matter configuration inside the star and also on the boundary vacuum conditions. 

The paper is organized as follows:
 In section \ref{sec_eqns} we formulate the theoretical framework for the gravitation. 
In section \ref{sec_sphsymm} we discuss the spherically symmetric spacetime and the objects classes.
In section \ref{sec_numerical} we describe how to derive the potential $V(\phi)$
from the field equations derived in \ref{sec_eqns} with the adequate boundary conditions and stellar 
configurations.
Finally, we draw our conclusions in section \ref{conclu}.

\section{The gravity formalism}\label{sec_eqns}
We consider a scalar-tensor action that is of Brans-Dicke type \cite{BD61} with a potential and
no kinetic term in the Jordan frame. 
This theory could also be equivalently described in the Einstein frame if also the units of length, 
mass and time are scaled accordingly. 
For this numerical work, however, the Jordan frame is preferred
\cite{0612075}.
The considered scalar-tensor action \cite{SotFar} is
\be{BDaction}
S=\frac{1}{2\kappa}\int d^4 x\sqrt{-g}\left[\phi R
-V(\phi)\right]+S_{\textrm{m}}, 
\ee
where $\kappa\equiv 8\pi G$, the $S_{\textrm{m}}$ term gives the matter contribution
and the scalar field $\phi$ interacts non-minimally with the gravitational field.
We use $c=\hbar=1$ in this work. This type of action is also equivalent to metric $f(R)$ theories of gravity
\cite{SotFar}.

The configuration is required to be GR$+\Lambda$ inside the matter configuration, to pass
the local gravity tests \cite{cassini,GRSStests} for the solar model.
The interaction with the matter is realized with the potential $V(\phi)$
and through the coupling term $\phi R$ in the used Lagrangian.
The field equations are derived from the action by varying with respect to the metric $g_\mn$ and 
also with respect to the field $\phi$, 
respectively:
\bea{BDfeqs}
G_{\mn}&=&\frac{\kappa}{\phi}T_{\mn}-\frac{1}{2\phi}g_{\mn}V(\phi)\nonumber\\
& &+\frac{1}{\phi}\left(\nabla_\mu\nabla_\nu\phi-g_{\mn}\Box\phi\right)\\
R &=&V'(\phi).\nonumber
\eea

The energy momentum tensor that describes the matter content is of standard perfect fluid form
$T=T^{\mu}_{\mu}=-\rho+3p$ \cite{MTW} and we use the polytropic equations of state 
to describe the adiabatic processes inside the star.
The matter term in (\ref{BDfeqs}) needs to obey the equation of continuity  
$D_\mu T^{\mu\nu}=0$, so this equation is used when the polytropic profile is calculated. 
The only non-trivial component of the equation of continuity for a spherically symmetric system  is 
\be{cont} 
p'=-\frac{B'}{2 B}(\rho+p).
\ee 
Here comma stands for the radial derivative, $'\equiv d/dr$.  
If the scalar-tensor gravity is to resemble general relativity at small scales, it should obtain the 
Schwarzschild solution outside a spherically symmetric object 
or the Schwarzschild-de Sitter vacuum outside the GR$+\Lambda$ star.
Since the potential $V(r)$ and the field $\phi(r)$ are compared to polytropes with observational properties, 
we study the field equations in the Jordan frame throughout this work for numerical convenience.

In the chameleon theories of gravity \cite{chameleon},
the effective potential is chosen such that the second derivative of the potential 
is dependent on the local energy density. In this way the fifth force is evaded with a 
sufficiently high mass of the field in dense environments.
In this work no fixed potential is selected, but the potential is 
required to stay close to the GR$+\Lambda$ solution inside the objects and to arrive at the
SdS vacuum solution. As a result the potential turns out to be very dependent on the
matter density and the boundary conditions.
Irrespective of whether the screening condition \cite{Waterhouse06} is fulfilled or not, a general potential 
$V(\phi)$ should be able to describe all the observed objects.
With the above conditions, every studied object class occupies an unique range for both $V(r)$ and $\phi(r)$,
namely $V(r,\phi(r))_{NS}\leq V(r,\phi(r))_{WD}\leq V(r,\phi(r))_{SUN}$.
Therefore, all the studied objects (SUN, WD, NS) cannot be described by one potential $V(\phi)$.

\subsection{Competence with the GR$+\Lambda$ and $f(R)$ models}
We demand the studied object to be comparable to an already established general relativistic configurations.
We, therefore, select for the scalar-tensor examination only the configurations
that are  similar enough to the general relativistic polytropes inside the star.
We study all the configurations separately with the GR$+\Lambda$ and the scalar-tensor
field equations, integrating first from the center outwards with general relativistic field equations and
then from the fixed SdS boundary inwards with the scalar-tensor field equations.
We also, for comparison, study configurations with $f(R)$ gravity interiors.

The  gravitational interaction that fixes the interior of the configurations is derived from the action
\be{fRaction}
S=\frac{1}{2\kappa}\int d^4 x\sqrt{-g} f(R)+S_{\textrm{m}}.
\ee
Now the Einstein-Hilbert action  that gives the GR$+\Lambda$ model is obtained with  $f(R)=R-2\Lambda$.

To numerically derive the polytropic profiles $(\rho(r), B(r), A(r))$ inside the configuration
we use  the temporal field equation from
\be{fRfeqs}
F(R) R_{\mn}-\frac 12 f(R) g_{\mn}-(\nabla_\mu\nabla_\nu -g_{\mn}\Box ) 
F(R)=8 \pi G T_{\mn} 
\ee
(with $\mu=\nu=0$).
The scalar degree of freedom is denoted here as $F(R)\equiv f'(R)$.
After obtaining $\rho(r),B(r)$ and $A(r)$, the scalar tensor potential $V(r)$ and field $\phi(r)$
are numerically integrated inwards starting from the surface.

Metric $f(R)$ theories, with an algebraic function $f(R)$ that replaces 
the Ricci scalar $R$ in the Einstein-Hilbert Lagrangian,  can be represented with the
scalar-tensor gravity of (\ref{BDaction})
\cite{scal-tens,SotFar}.
In this work we, therefore, constructed the potential $V(\phi)$ and the field $\phi$ 
also by demanding that the polytropes'
energy density and the metric follow the field equations for two $f(R)$ models.
We in particular studied the chameleon $f(R)$ gravity models of Hu and Sawicki (HS) and Starobinsky 
(St), that can produce the correct cosmological dynamics and have the correct weak field limit.
The $f(R)$ function for the Hu-Sawicki model \cite{HS} is 
\be{fRHS} 
f(R)_{HS}=R-m^2\frac{c_1(\frac{R}{m^2})^\alpha}{c_2(\frac{R}{m^2})^\alpha+1},
\ee
where $\alpha >0$, $c_1$ and $c_2$ are dimensionless parameters and $m^2$ is the mass scale of the vacuum today.
The Starobinsky model \cite{Starob} is
\be{fRSt}
f(R)_{St}=R+\lambda R_\Lambda \left( \left(1+\frac{R^2}{R_\Lambda^2}\right)^{-\beta}-1\right),
\ee
where $\beta,\lambda >0$ and $R_\Lambda$ is the vacuum scale.
The field equations are highly non-linear, and for the functions (\ref{fRHS}) and (\ref{fRSt}),
therefore, only solar type polytropes and white dwarfs are solvable with the current code.

We also separately  solve  the Tolman-Oppenheimer-Volkov (TOV) equations \cite{WEI} 
for all the selected objects
to make sure  the objects don't deviate much from GR \cite{MV0612775}. 
All the chosen polytropes (also the ones derived with the $f(R)_{HS,St}$ field equations) 
separately fulfill both (GR+$\Lambda$/$f(R)$ and TOV) requirements.

\section{Static spherically symmetric solutions} \label{sec_sphsymm}
We consider here static, spherically symmetric bodies embedded in background a SdS vacuum
with $R_0=12 H_0^2$  \cite{0610296}.
We use the general spherically symmetric line element
((8.1.4) in \cite{WEI})
\be{metric}
ds^2=B(r)dt^2-A(r)dr^2-r^2(d\Theta^2-\sin^2\Theta\, d\Phi^2),
\ee
Where $A(r)$ and $B(r)$ are free functions that are numerically derived inside the stellar body from
the GR$+\Lambda$ field equations with the polytropic EOS.
The angle coordinates $\Theta$ and $\Phi$ in (\ref{metric}) 
do not enter the used equation ansatz at all. 

The polytropes are fixed to the SdS vacuum at the boundary \cite{HMV07} to make sure 
the initial values $V_i, \phi_i, \phi_i'$ correspond to stellar objects with a 
cosmological vacuum.
The scalar-tensor field equations are
integrated from the surface to the center with the scalar tensor field equations (\ref{BDfeqs})
starting from the Schwarzschild-de Sitter metric 
\bea{SdSmetric}
B(r)=c^2 \left(1- \f{2 G M}{c^2 r_s} - \f{\Lambda}{3} r_s^2\right),\nonumber\\
A(r)=\left(1- \f{2 G M}{c^2 r_s} - \f{\Lambda}{3} r_s^2\right)^{-1}
\eea
at the boundary.
Here the mass $M$ and the stellar radius $r_s$ are obtained first with the GR$+\Lambda$ field equations.

The scalar-tensor field equations (\ref{BDfeqs}) can be analytically solved in the 
de Sitter vacuum $$G_{11}=-3A(r) H_0^2,\quad T_{11}=0$$ and the first field equation
boils down to
\be{SdSeqs}
V(\phi)=-\left(\frac{B'}{B A}+\frac{4}{Ar}\right)\phi' +6H_0^2\phi
\ee
at the surface with
$R=4\Lambda\equiv 12 H_0^2$  \cite{0610296}.
The Schwarzschild-de Sitter vacuum can be obtained for $V=2\Lambda=6H_0^2$, 
$\phi=1 $ and $ \phi'=0$, but in this numerical analysis
$V=6H_0^2$ is obtained with the field values deviating from these
a little $\phi\approx 1 $ and $ \phi'\approx 0$.
The potential outside the body stays near the vacuum value, but the field and its derivative develop.
The field behavior outside is dependent on the stellar boundary conditions
and the field value will be a monotonically decreasing function of the radius.
With identical initial $\phi_i(r_s)$ at the boundary, physically valid configurations can be found, but
the potential magnitude $V(r)$ inside will still be dictated by the matter configuration and
will range different values for each object class.

\subsection{Polytropic stellar configurations}\label{sec_polytr} 

Polytropic model describes adiabatic processes. 
Stars are often modelled with the polytropic model because it naturally results in a
monotonically decreasing density profile with a well defined boundary \cite{Chandrasekhar}.
The polytropic equation of state (EOS) is
\be{eos}
p(r)= K \rho(r)^\gamma \ee
Here $K$ is a constant, and $\gamma$ is related to the polytropic index $n$ by
$\gamma=n/(1+n)$.
This restricts the perfect fluid matter to form spherical objects that are most dense at the core.
All the configurations we studied here are tested to be regular at the center \cite{HMV07}, having 
physical central densities $\rho_c$ and obtaining masses and radii that conform to observations.
We have considered only objects that are also numerically equivalent to their 
Tolman-Oppenheimer-Volkov counterparts \cite{WEI} by separately solving the TOV equation for the 
same stellar configurations and by comparing the solutions.

The parameter space ($K, \rho_c, \gamma$), yielding the physically and observationally acceptable
object classes is extremely tight for compact configurations. However, one can find objects that represent
relativistic and non-relativistic white dwarfs as well as neutron stars that also follow
the $R\sim \kappa\rho$  behavior inside the body. 
One representative from each class was chosen for plotting the scalar field potential
$V(\phi)$ in Fig(\ref{NSWDSUN}).

We use the Eddington polytropic model of \cite{HV14} with $n=3$ to produce a representative of the
stellar object class.
The two other studied classes of compact stars are
polytropic white dwarfs,
and polytropic neutron stars.
The polytropes'  matter density is parametrized as in the general relativistic Lane-Emden case \cite{WEI}
with $\rho(r)=\rho_c\theta^n(r)$ for the numerical work ($\theta$ is the scaled density parameter). 
The Lane-Emden equation itself is not used in this work, only the above parametrization.

Stability considerations for the scalar-tensor polytropic configurations have not been examined 
in this work. However we list here some works done with general relativistic polytropes.
Here, we present some aspects to take into account when considering the stability of
static polytropes also with modified gravity.
All static polytropic solutions in GR are considered stable in  \cite{Fronsdal2008} 
if the solution is regular 
at the center and the density falls of rapidly after the boundary region. 
Also note that due to the higher order nature of the 
solutions for higher order gravity theories, 
the boundary matching at the surface is not discontinuous and SdS can be reached naturally.
Furthermore, for static spherically symmetric solutions in scalar-tensor gravity, the presence of 
a non-negative effective potential implies the absence of unstable modes for 
linear perturbations in the scalar field \cite{Harada1997}.
Kosambi-Cartan-Chern and Lyapunov stability properties of Lane-Emden equations have been studied
in \cite{polystab}. With both these methods the general relativistic polytropic index is  stable only for 
the values
$\gamma \in[1.2,1.313708]$. This does not, however, include the solar Eddington polytrope with 
$\gamma_{SUN}=4/3$ nor the non-relativistic white dwarfs with $\gamma_{WDnr}=5/3$ although these polytropic 
equations of state are widely used for modelling polytropic stellar type stars and white dwarfs.

Polytropic stellar models are not realistic, but extremely useful for their simplicity and fairly good 
resemblance to more accurate models in the stellar case. Also, 
polytropic stars arise naturally in general relativity \cite{WEI,Chandrasekhar} and provide
a good approximation for the observed systems.
We are using GR$+\Lambda$ equations inside these objects, because
the Birkhoff's theorem conforms to the Einstein-Hilbert action with $\Lambda$.
These objects should, therefore, be valid also for models in the modified gravity sector.

In this work we discuss observationally and physically acceptable self-consistent objects 
in the studied scalar-tensor gravitation. The configurations are demanded to verify the following 
requirements:
\begin{itemize}
\item[i)] The studied polytropes were chosen from the object class such that 
they obtain  masses and radii within observed ranges. Also, the central densities 
need to be of the right order wrt solar standard model \cite{Bahcall04} and Harrison-Wheeler EOS \cite{MTW}.
The parameter values are discussed in the following subsections.

\item[ii)] The parameters $\rho(r), A(r)$ and $B(r)$ are first derived with the GR$+\Lambda$ or $f(R)_{HS,St}$ 
field equations, which selects a scalar-tensor models with a potential $V(\phi)$ that conforms to
general relativity inside the object \cite{MV0612775}.

\item[iii)] We demand the spacetime to conform to SdS at the boundary by numerically solving (\ref{SdSeqs})
with $V=6H_0^2$ and the SdS metric parameters $B(r_s)$ and $A(r_s)$ in (\ref{SdSmetric}), that were obtained 
with the mass $M$ and the radius $r_s$ of a GR$+\Lambda$ polytrope.

\end{itemize}

\subsubsection{Solar and white dwarf polytropes}
For the stellar type stars, we use the Eddington model that was also studies in context of Hu-Sawicki $f(R)$
gravity in \cite{HV14}. Eddington model gives a fair approximation for the 
standard solar model density profile \cite{Bahcall04} with $n=3$ and half the standard model central density.

We studied two classes of WDs with different polytropic indices,
$n=1.5$ for non-relativistic and $n=3$ for relativistic white dwarfs.

The observational mass and radius ranges for the WDs were referenced from observations; 
see \textit{e.g.} \cite{WDobs1}.
We accepted as valid values for the mass $\in[0.4,0.7]\,M_\odot$ and 
for the radius  $\in[8000,11000]\,$km.
The coefficient for the polytropic equation of state for non-relativistic degenerate gas is $n=1.5$
and for highly relativistic degenerate gas $n=3$.
Observational masses and radii for the white dwarfs
were found only for the physical central densities that conform to the 
Harrison-Wakano-Wheeler stellar models (\cite{MTW} p.625).

\subsubsection{Polytropic neutron stars}
There are many rival models for the structure of a neutron star and there is no 
favorite equation of state to be used.
Many modern models build the neutron star from multiple polytropic layers as well as 
separate crust or core with different equation of state \cite{NSth1,NSth2,Lattimer13}.
In this work will use two layers of polytropes of which the first describes the core area $\in[r_0,r_c]$ 
(about 10\% of the radius of the NS) and the second layer extends from the core to the surface $r_s$.

The observational mass and radius ranges we use conform to \cite{Lattimer13,NSobs}
As valid  mass range we accepted the observational values $\in[1.4,1.7]\,M_\odot$ and the corresponding
radii estimates (that are in accordance with the studies \cite{NSth1,NSth2,DH2001}) 
to lie in the range $[9,13]$ km.

For the HS and St field equations, neutron stars do not solve even for the singly-polytropic 
object with the current code.

\section{Numerical work}\label{sec_numerical} 
One observationally valid object (with particular $K,\rho_i$ and $\gamma$) was chosen to represent 
each stellar class. Inside the star, the GR$+\Lambda$ field
equations  were solved starting from a smooth center to obtain
$\rho(r), A(r)$ and $B(r)$. 
The scalar-tensor potential $V(r)$ and the field $\phi(r)$ inside the body were obtained with the
equations (\ref{BDeqns}) inside, starting from the boundary with the metric initial conditions
set to SdS (\ref{SdSmetric}).

Also the field equations corresponding to $f(R)$ gravities 
(\ref{fRHS}) and (\ref{fRSt}) were used to obtain the comparison $f(R)$ configurations.
These configurations are only shortly discussed in this text and the focus is on the  GR$+\Lambda$ 
matter configurations.

To constrain the parameters in the scalar-tensor field equations (\ref{BDfeqs}),
we bind together the metric parameters, pressure and the matter density inside the configurations
by demanding the  GR$+\Lambda$ field equations (from (\ref{fRaction}) with $f(R)=R-2\Lambda$), 
the continuity equation (\ref{cont})  and the 
polytropic equation of state $p=p(\rho)$ to hold for each object class. 
The used equation ansatz  used to derive the stellar interior consists of the $\mu=\nu=0$ 
-field equations (\ref{fRfeqs}),
the trace equation
\be{fRtrace}
F(R)R - 2 f(R) + 3\Box F(R) = 8 \pi G (\rho - 3 p), 
\ee
the Ricci scalar as a function of the metric parameters $R[B(r),A(r)]$ and 
the continuity equation (\ref{cont}).

The curvature scalar $R$ follows the energy density $\kappa\rho$ for all the selected WDs and SUN
and the solution was checked to correspond to the general relativistic counterpart 
that was solved separately for the same polytropic parameters with the TOV equations \cite{WEI}.
The TOV equation was built as a separate code and used only for comparison  to
judge if the density $\rho(r)$ in the modified code obtains the general relativistic TOV $\rho_{TOV}(r)$ solution
with high accuracy.

The polytropic coefficient $K$ was chosen such that mass
and radius, that conform to the observations, are produced when the object is integrated from inside out with
the GR$+\Lambda$ field equations. A wide range of polytropic coefficients $K$ 
and central densities, $\rho_c\in[10^4,10^{11}]\, \textrm{kg m}^{-3}$, in (\ref{eos})
 were also scanned to be sure no other region of solutions for 
observationally and physically valid solutions exist.
The domain of observationally and physically
valid polytropic solutions shrinks essentially to a point in the parameter space.

The potential $V(r)$ and the field $\phi(r)$ were solved numerically from the contraction of the 
11-component of the scalar-tensor field equations (\ref{BDfeqs}) with the metric (\ref{metric})
\bea{BDeqns}
V(r)&=&2\kappa p(r)(1-\phi(r))-\frac{4 +\frac{B'(r)r}{B(r)}}{A(r)r}\phi'(r)\nonumber\\
V'(r)&=&\phi'(r)R(r).
\eea 
Note that we have reparametrized the potential $V(\phi)\rightarrow V(r)$, thus the form of the second 
field equation is  changed.
The integration of $V(r)$ and $\phi(r)$ was started from the surface inwards with the 
initial conditions chosen by the SdS metric vacuum requirement (\ref{SdSeqs}).

The dimensionless field $\phi$ must be of order unity \cite{SotFar} 
to be consistent with the solar system experiments. The field naturally varies extremely little from unity
with the SdS condition at the boundary; 
only  $\mathcal{O}(\delta\phi)\sim 10^{-17}$ for the solar class, SUN, and less for the WDs and NSs. 
Therefore, the effective
parameter in this work is the deviation of the field $\phi=1+\delta\phi$.
The initial condition $V_i=V(r_{s})$ was set to be the SdS vacuum and $\phi_i'$
is solved from (\ref{SdSeqs}) by using 
$\phi_i=\phi(r_{s})$  as a free parameter at the surface of the star.
The sensitive boundary conditions  for $\delta\phi_i$ range from $10^{-20}$ to $10^{-29}$ for the SUN and NS
case respectively, see Fig.(\ref{NSWDSUN}).

\subsection{Results}\label{sec_BD}
The results presented here are qualitative in their nature although specific numerical values 
play a crucial role in the numerical analysis.
In this article we have derived the scalar field potential $V(\phi)$ of the action (\ref{BDaction})
for polytropic stellar configurations that conform to general relativistic stars (solar,
white dwarf and neutron star polytropes).
As the main result, we consider the dependence of the numerically derived scalar field potential $V(\phi)$ 
on the matter configuration inside the stellar body in each case. This property does not allow to describe
all the studied stellar objects with the same potential (that is with the same theory).
We also find, that the resulting field inside the configuration $\delta\phi(r)$ is highly dependent on the 
initial value $\delta\phi_i$ at the surface.

\begin{figure}[t!] 
\begin{center} 
\includegraphics[width=0.5\textwidth,angle=0]{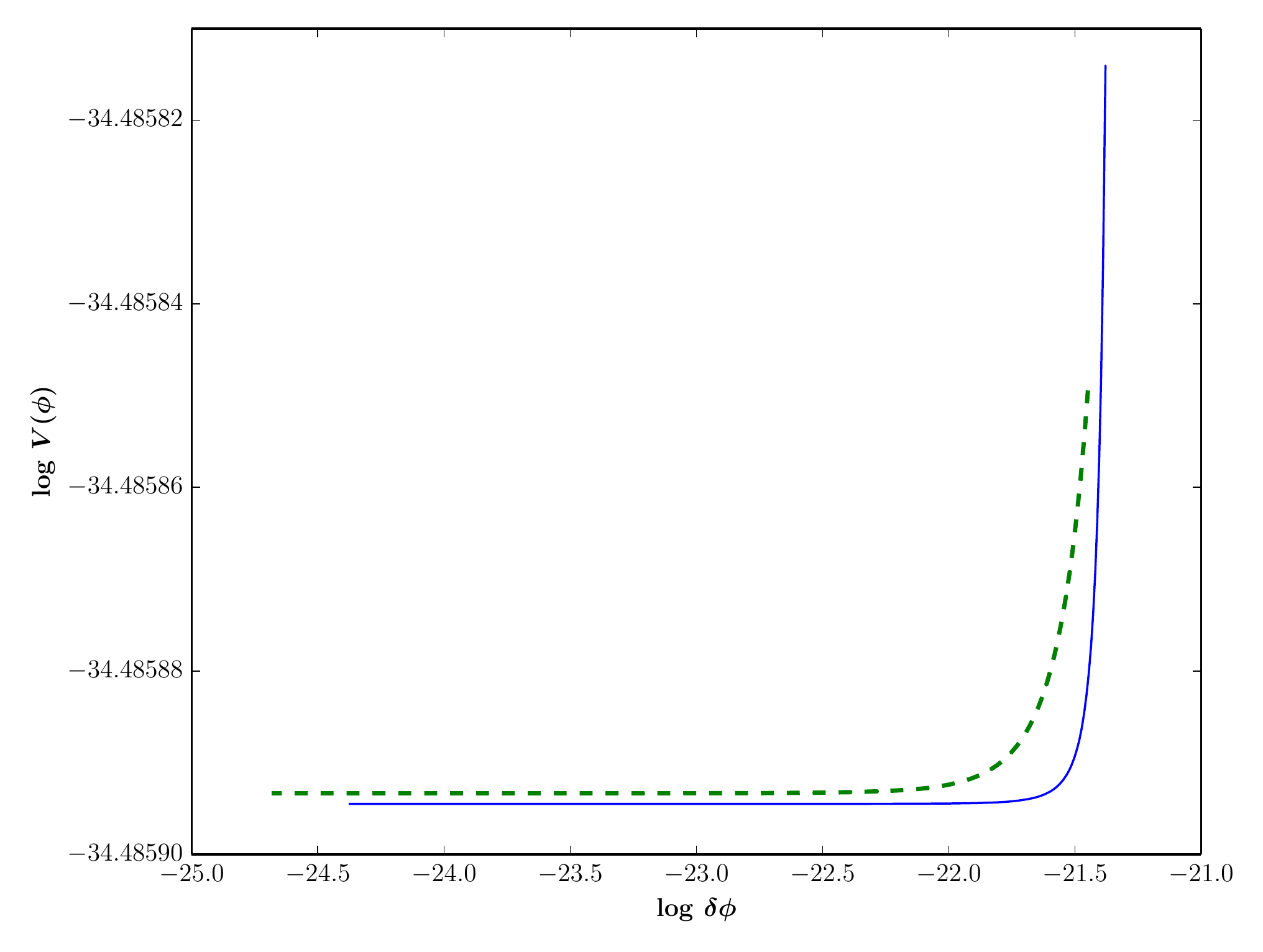} 
\caption{(Color online)
As an example a non-relativistic (green/dashed)  and a relativistic (blue/solid curve) white dwarf, 
showing the typical field and potential behavior inside the configuration. 
The potential $V(\phi)$ is highly dependent on the matter density inside the object.
The initial value at the surface was chosen to be the first $\phi_i'$ value that produces the SdS vacuum 
solution (\ref{SdSeqs}) with the potential value $V_i=2\Lambda=6H_0^2=3.2667\times 10^{-35} s^{-1}$.
}
\label{PotPhiWD} 
\end{center} 
\end{figure}

One can fix the higher order terms of the metric parameters \cite{SotFar} at the boundary,
therefore, the transition
from the polytropic configuration to the outside solution at the surface $r_s$ is smooth.
The potential and the field are monotonically increasing functions of the radius from the surface inwards
and will produce $V(\phi)$ that is monotonically increasing in the positive field direction 
Fig.(\ref{PotPhiWD}). 
The potential values in the plots for the examined stellar objects  ranges from the surface values, 
$V_i$, outside up to a  value  depending on the density at the core.
Because of this,  a neutron star will always
reach  higher potential values than a white dwarf or a solar type main sequence star
$V(\phi)_{NS}\leq V(\phi)_{WD}\leq V(\phi)_{SUN}$.

\begin{figure}[t!] 
\begin{center} 
\includegraphics[width=0.5\textwidth,angle=0]{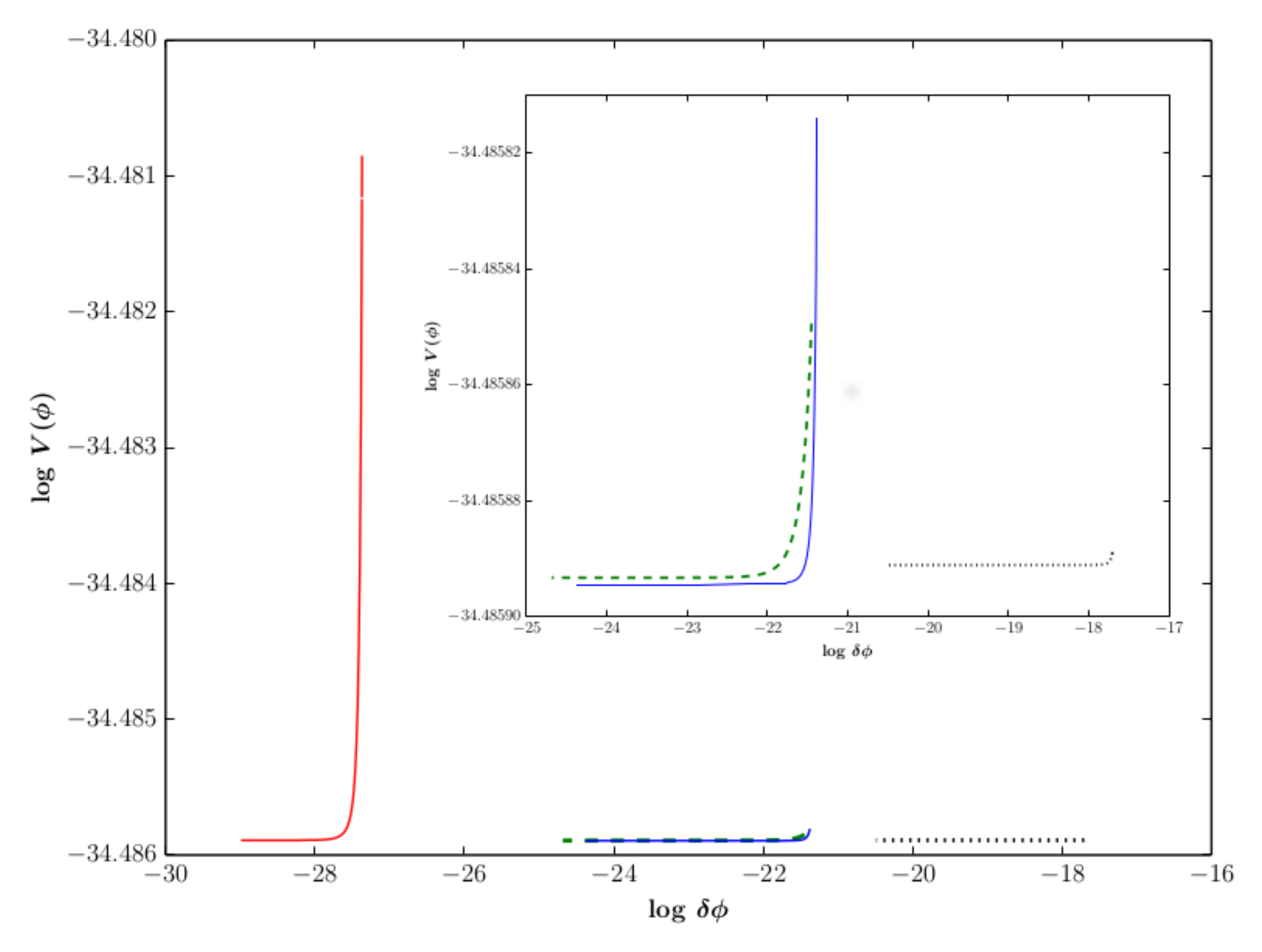} 
\caption{
Here the non-relativistic neutron star (red, leftmost), white dwarfs (blue and green/dashed, center) 
and a Sun like star (black/dotted, right) all have 
de Sitter vacuum as the surface initial condition (at $log_{10}(V(\phi_i))=-34.485891$). 
The smaller figure is a zoom in to the WDs' and the SUN's region.
All the objects show similar dependence with respect to the matter density. That is, all the
potential curves are monotonically increasing functions of the radius, increasing toward the center. 
}
\label{NSWDSUN} 
\end{center} 
\end{figure}

All the objects in the figures Fig.(\ref{PotPhiWD},\ref{NSWDSUN})  
in principle  reach the SdS with $V=6H_0^2$.
The plots, however, will only show data down to a finite field value due to discrete numerical methods.
Also, there is a limiting value
for the $\delta\phi_i$ that {\it saturates} the field to a minimum value that depends on the object.
This will define a typical field range for each object class that is shown for all the studied classes on 
the Fig.(\ref{NSWDSUN}).

We will consider the solar case, SUN, as a general relativistic polytrope that conforms to the 
Cassini results \cite{cassini}.
The initial condition $\delta\phi_i$ can be relaxed from the saturated case such that the WDs and the NS
obtain solutions that lie within the same $\delta\phi$ range. However, the field value ranges correspond 
to the size of the object, so that the more compact objects will always obtain smaller ranges than the 
more extended cases. 
The potential values, being dependent on the density and not on the field initial value, 
will in any case be higher for the denser objects, therefore, a common function $V(\phi)$ cannot be found.
Considering this,  a potential that matches  all the polytropic
stellar configurations and also conform to observations could be found, even with considerable fine-tuning.

The polytropic profiles and the metric parameters calculated from the Hu-Sawicki and 
Starobinsky field equations 
yield solutions for the potential $V(r)$ and the field $\phi(r)$ that are similar to the 
GR$+\Lambda$ solutions. 
Almost identical solutions can be found for all the  objects except the NSs that could not be solved
for the $f(R)$ field equations.

\section{Conclusions and discussion}\label{conclu}
We considered configurations of general spherically symmetric, static spacetimes 
with adiabatic perfect fluid matter in the scalar-tensor gravity (\ref{BDfeqs}). 
These  solutions were required to follow general relativity 
with $\Lambda$,  according to the Einstein-Hilbert action,
or $f(R)$ gravity with (\ref{fRHS}) and (\ref{fRSt}) inside the polytropic configurations.
The field equations (\ref{BDfeqs}) were numerically solved to
arrive at the scalar-tensor potential $V(\phi)$ inside the stellar object.
We studied examples of polytropic stellar object classes; a solar type star, non-relativistic and 
relativistic white dwarfs and a neutron star with observationally acceptable parameters.

We first solve the field equations numerically
for the metric functions $A(r)$ and $B(r)$
and for the energy density $\rho(r)$ inside the polytropic object with (\ref{fRfeqs}) 
starting from a smooth center.
Then  obtain the field $\phi(r)$ and the potential $V(r)$  with (\ref{BDfeqs}) 
starting the iteration from
the surface inwards by choosing the boundary conditions to accept the Schwarzschild-de Sitter solution with
 $V=6H_0^2$.

As a result, we find that the potential of a scalar-tensor polytrope that conforms to general relativity
is highly dependent on the matter configuration. Also, the possible field values are defined by the vacuum 
initial conditions down to a minimum value that is unique to the stellar object class. 
A potential $V(\phi)$ that would correspond to all the polytropic objects could not be found.
The stability with respect to the polytropic equation of state in scalar-tensor gravity has not been 
studied yet, so it is possible that some of the objects described here are not stable.

\acknowledgments 
K.H. would like to thank V.~Faraoni for useful comments. K.H was funded by the
University of Turku, UTUGS grant.

\end{document}